\documentclass[12pt]{article}
\usepackage{epsfig}
\begin{document}

\begin {center}
{\Large { A PHYSICAL INTERPRETATION OF  MOND}}
\vskip 5mm
{\rm D.V. Bugg \footnote {david.bugg@stfc.ac.uk}} \\
{\normalsize  Department of Physics, Queen Mary, University of London, 
London E1\,4NS, UK} \\
[3mm]
\end {center}

\begin{abstract}
\noindent
Earlier comparisons of galatic rotation curves with MOND have arrived at the
conclusion that the parameter $a_0$ lies within $\sim 20\%$ of  $cH_0/2\pi$, 
where $c$ is the velocity of light and $H_0$ is the Hubble constant.
It is proposed here that,  for this value of $H_0$, signals propagating around 
the periphery of the Universe are phase locked by the graviton-nucleon 
interaction.
\end {abstract}

\section {Introduction}
MOND (Modified Newtonian Dynamics), is an empirical scheme invented by Milgrom
in 1983 \cite {MilgromA}, \cite {MilgromB} to account for what was then known 
about rotation curves of galaxies from the work of Tully and Fisher \cite {Tully}.
Famaey and McGaugh have reviewed all aspects of the data with an exhaustive list 
of references \cite {Famaey}.
A valuable review of MOND and its foundations has appeared in a recent article by 
Sanders \cite {Sanders}.

In MOND, the observed total acceleration $A$ is related to Newtonian 
acceleration $g_N$ by
\begin {equation}     %Eqs, 1
A = g_N/ \mu (\chi)
\end {equation}
where $\mu$ is a smooth fitting function, $\chi = A/a_0$ and $a_0$ is a 
universal constant with a value $\sim 1.2 \times 10^{-10}$ m s$^{-2}$ for all 
galaxies.
There are three forms in common use.
All three tend to $A = \sqrt {a_0g_N}$ for small $g_N$.
The most symmetrical is 
\begin {equation}      %Eq. 2
\mu = \sqrt {1 + \frac {a_0^2}{4A^2}} - \frac {a_0}{2A}.
\end {equation}
Re-arranging this expression and equation (1)
\begin {eqnarray}     %Eqs. 3-4
(g_N/A + a_0/2A)^2  & = & 1 + (a_0/2A)^2 \\
g^2_N + a_0g_N & = & A^2.
\end {eqnarray}
A star with rotational velocity $v$ in equilibrium with centrifugal force obeys
\begin {equation}        %Eq. 7
v^2/r = \sqrt {a_0GM}/r;
\end {equation}
 $G$ is the gravitational constant and $M$ the galactic mass within radius $r$; 
the factor $r$ cancels  and the Tully-Fisher relation emerges: 
\begin {equation}     %Eq. 8
v^4 = a_0GM.
\end {equation}
Famaey and McGaugh show examples comparing this relation with rotation curves in 
their Figs. 24 to 27. 
McGaugh demonstrated that galaxies with well determined rotational velocities 
follow the Tully-Fisher relation accurately over 5 decades of galactic mass 
\cite {McGaugh}.

In a later paper, McGaugh concludes that gas-rich galaxies give the best 
determination of baryonic masses of galaxies, with the result 
$a_0 = (1.113 \pm 0.3) \times 10^{-10}$ m s$^{-2}$ \cite {McGaugh2}. 
Gentile, Famaey and de Blok  found $1.122 \pm 0.33 \times 10^{-10}$ m s$^{-2}$ 
\cite {Gentile}.

\section {The relation of $a_0$ to the Hubble Constant}
The radius of the Universe is given by the Hubble length $r = c/H_0$.
At the velocity of light, signals propagating round  the periphery of the 
Universe in any direction take time $2\pi c/H_0$ to complete one orbit.
The parameter $H_0$ has the dimensions $s^{-1}$, with the result that 
$2\pi c/H_0$ has the dimensions of acceleration, as does $a_0$. 
Gravitational waves propagating round the periphery of a spherical  Universe 
in all directions then arrive in phase.
Fig. 1 shows the observed variation in galaxies between the total acceleration 
$A$ and Newtonian acceleration $g$. 
The conclusion was reached in an earlier paper \cite {Bugg} that galaxies are 
Fermi-Dirac condensates in the graviton-nucleon interaction.
Consider  gravitons arriving at the edge of a galaxy and distributed around an 
axis running from the periphery of the galaxy to its centre. 
These gravitons have very long wave-lengths and can interfere coherently with 
stars and clusters of stars over a very large volume. 
The overall result is given by the $\it {intensity}$ of this interaction 
acting as an amplifier.
The result may be fitted by an Airy disc with an adjustable radius parameter.
The shape of Fig. 1 defines the Airy disc.

As aside is that there was one unfortunate error of wording in Ref. 
\cite {Bugg}. 
The text referred to the Hubble acceleration; this should read Hubble constant. 
It has no effect on conclusions in that paper.

%Fig. 1
\begin{figure}[htb]
\begin{center}
\vskip -8mm 
\epsfig{file=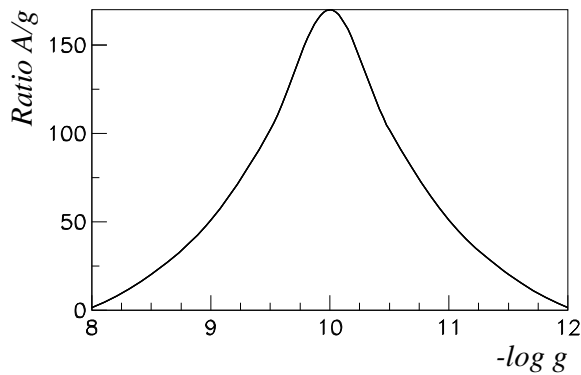,width=10cm}
\vskip -4mm
\caption
{ The observed ratio between total acceleration $A$ and Newtonian 
acceleration $g$.}
\end{center}
\end{figure}

My proposal is that the same mechanism operates in the graviton-nucleon 
interaction on the scale of the Universe. 
There are inhomogeneities in the Universe, notably the Sloan Great Wall. 
This will induce a similar shape to that in Fig. 1. 
We can only see such features as they were a long time ago.
They will change with time as galaxies evolve and crash into one another.
The fact that the Universe is round suggests that effects of this evolution 
are damped oscillations; it does not go pear-shaped.
The inference is that the parameter $a_0$ originates from the curvature of 
space.
Milgrom forsaw this in the second of his early papers on MOND \cite {MilgromB}. 
If $H_0$ changes, my prediction is that $a_0$ will change with it.

\section {An important comment on how to fit the Cosmic Microwave 
Background}
There is an important result from a paper of Milgrom on  weak gravitational 
lensing of galaxies \cite {MilgromW} using data of Brimioulle et al. 
\cite {Brimioulle}.
They have examined foreground galaxies illuminated by a diffuse background 
of distant galaxies.
They removed signals from the centres of foreground galaxies so that their 
edges and haloes could be studied.
Their objective was to study the Dark Matter halo.
The asymptotic form $\sqrt {a_0 g}$ of the acceleration varies as $1/r$;
integrating this term leads to a logarithmic tail 
$V(r) = -\sqrt {GMa_0} \ln _e(r/r_0)$ to the Newtonian potential; here 
$r_0$ is the mean radius for this term.
Milgrom transforms this into the variables used by Brimioulle et al. 
He shows that their results obey MOND predictions accurately over a range 
of accelerations $10^{-9}\, -\, 10^{-11}$\, m \, $s^{-2}$.
Averaged over this range, results are a factor $\sim 40$ 
larger than predicted by conventional Dark Matter haloes surrounding 
galaxies.
Fig. 1 shows the ratio of observed acceleration $A$ to Newtonian 
acceleration as a function of $x = -{\rm {\log }_{10}} \, g_N$.  
At the peak acceleration $a_0$, the effect is larger than Dark Matter 
predicts by a factor $\sim 65$.
The conclusion is that the standard $\Lambda CDM$ model needs serious 
modification.

There are two sources of information on $H_0$ and its variation with time.
One is the observed variation of type 1a supernovae as a function of 
red-shift $z$.
The second is the map of temperature fluctuations observed over the 
visible Universe.
This spectrum is expressed as a series of Spherical Harmonics up to angular 
momentum $L \simeq 2500$.
Peaks appear in the spectrum beginning at $L \simeq 220$ and falling as 
$L \to 2500$.
A critical issue is how to interpret this spectrum.

The recombination of hydrogen atoms after the Big Bang occurred over a 
volume with a diffuse edge.
The slope of this edge governs the magnitudes of successive peaks 
between $L=220$ and 2500.
It arises from damping of the power spectrum at high $L$ values due to 
photon diffusion from high temperature regions to cooler regions, Silk 
damping \cite {Silk}.
In the correlations of temperature $T$ in the data published recently 
by the Planck collaboration \cite {PlanckH0}, there is a contribution 
from correlations $TT$ in the Baryon Acoustic Oscillations and also 
E- and B-mode polarisations.
Formulae are given in Section 8.7 of a paper by Straumann \cite {Straumann}
and have been used by the Planck collaboration.
A recent paper of Schmittfull, Challenor, Hanson and Lewis \cite {Schmittful} 
reviews in detail the fit to data.

However, there is a serious point which has been overlooked.
A corollary follows from Milgrom's fit to the data of Brimioulle et al. which 
agree with MOND, but are far from the prediction of $\Lambda CDM$. 
Here the photons come from distant galaxies. 
The Baryonic Acoustic Oscillations are likewise carried by photons, which in 
this case originate from the Cosmic Microwave Background. 
These must be treated in the same way.
The conventional assumption made in the work of Ref. \cite {Schmittful} is 
that photons from the Cosmic Microwave Background are bent in weak 
gravitational lensing only by Newtonian dynamics (including small 
corrections for General Relativity).
However, since MOND fits the data of Brimioulle et al. but $\Lambda CDM$ does 
not by a large margin,  the astrophysics community should be alert to the 
fact that an additional energy $-\sqrt {GMa_0} \ln  _e(r/r_0)$ originates 
from integrating the acceleration $\sqrt {GMa_0}/r$;
$r_0$ is the radius where the acceleration is $a_0$.
This is needed over the range of accelerations where MOND explains the 
gravitational rotation curves, see Figure 1.
It is not presently included in the fit to the Baryonic Acoustic 
Oscillations but needs to be.

Since $\Lambda CDM$ disagrees with the data of Brimioulle et al., a new
paradigam is required, as argued in a somewhat different context by
Kroupa, Pawlovski and Milgrom \cite {Kroupa}.

It is now well established that there is a cosmic web of filaments, galaxies 
and voids on the scale of the Universe. 
These appear in WMAP data on very large scales, correponding to an angular 
dependence $L \leq10$ on the scale of the Universe \cite {WMAP}. 
Two recent papers on the cosmic web are by Alpaslan et al. \cite {Alpaslan} 
and Cantun et al. \cite {Cantun}. 
My proposal is that these structures again arise from the graviton-nucleon 
interaction on the scale of the Universe.

\section {Further Comments}
It is important to realise that the argument so far concerns the Hubble 
constant at a particular time.
It evolves with red-shift, as observed by the late-time acceleration of the 
expansion of the Universe.
This is conventionally described by the Friedmann-Lema\^ itre-Robertson-Walker 
metric currently used to fit Dark Energy. 
This is a separate issue.

Another point concerns the precise shape of Fig. 1. 
This was derived in Ref. \cite {Bugg} which gives values of the ratio $A/g$ 
at values of $x = -\log g$; these are 171 at $x = 10$, 85 at $x = 10.6$, 
30 at $x = 11$, 6 at $x = 11.5$ and $\sim 1$ at $x = 12$.
The curve is symmetrical about $x = 10$. 
However, there are uncertainties in these values up to $\sim 4\%$ of the 
peak value, i.e. $\pm \sim 2$, in the tail of the distribution for $x >11$.

There has recently been great interest in results of the BICEP 
collaboration \cite {Bicep}.
Astrophysics groups should be aware that these results also require the 
same correction for the additional energy $-\sqrt {GMa_0} \ln _e(r/r_0)$ 
when fitting the Cosmic Microwave Background.  

\section {Conclusions}
The primary conclusion is based on the fact that the parameter $a_0$ of 
MOND agrees with $cH_0/2\pi$.
It may be understood as arising from the graviton-nucleon interaction. 
It was proposed earlier \cite {Bugg} that galaxies are Fermi-Dirac 
condensates formed in this interaction.
Here it is proposed that the same graviton-nucleon interaction operates 
on the scale of the Universe and leads to phase-locking of signals 
propagating round the periphery of the Universe in any direction, hence
the shape of the curve in Fig. 1. 

A second conclusion is that the Cosmic Microwave Background needs to be 
fitted including a correction which derives from an additional energy 
$-\sqrt {GMa_0 \ln _e(r/r_0)}$ which is presently missing from equations 
for the Cosmic Microwave Background.
One should remember that there has so far been no significant evidence 
for the detection of Dark Matter particles above the level of $\sim 3$ 
standard deviations, although there is no shortage of ideas what it 
might be.  
The consequence is that the Cosmic Microwave Background is the primary 
source of the relative contributions to the Universe of Dark Matter and 
Dark Energy today. 
The Baryonic Oscillations are presently fitted with 6 parameters, 
sometimes 9. 
Any error in allowing for variation of peak heights in these oscillations 
potentially has large repercussions on conclusions on galaxies and the 
Universe as a whole.


\begin {thebibliography} {99}
\bibitem {MilgromA}    %1
M. Milgrom, 1983, ApJ {\bf270} 371.  
\bibitem {MilgromB}    %2
M. Milgrom, 1983, ApJ {\bf 270} 384. 
\bibitem {Tully}       %3
N.B. Tully and J.R. Fisher, Astron. Astrophysics {\bf 54} 661 (1977).
\bibitem {Famaey}      %4
B. Famaey and S.S. McGaugh arXiv: 1301.0623.
\bibitem {Sanders}     %5
R.H. Sanders arXiv: 1404.0531. 
\bibitem {McGaugh}     %6
S.S. McGaugh,  ApJ {\bf 632} 859 (2005). 
\bibitem {McGaugh2}    %7
S.S. McGaugh arXiv: 1107.2934.
\bibitem {Gentile}     %8
G. Gentile, B. Famaey and W.H.G. de Block Astron. Astrophys. {\bf 527} A76 
(2011).
\bibitem {Bugg}        %9
D.V. Bugg, Can. J. Phys. {\bf 91} 668 (2013), arXiv: 1304.7483.
\bibitem {MilgromW}    %10
M. Milgrom, Phys. Rev. Let. {\bf 111} 041105 (2013).
\bibitem {Brimioulle}  %11
F. Brimioulle, S. Seitz, M. Lerchster, R. Bender and R. Snigula, 
arXiv: 1303.6287. 
\bibitem {Silk}        %12 
J. Silk, ApJ, {\bf 151}, 149 (1968). 
\bibitem {PlanckH0}    %13
P.A.R. Ade et al. (Planck Collaboration), arXiv: 1303.5075.
\bibitem {Straumann}   %14
N. Straumann, Annal. Phys. {\bf 15}, 701 (2006)  and arXiv: hep-ph/0505249.
% the arxiv reference may ease the problem of locating Annal. Phys.
\bibitem {Schmittful}  %15
M.M. Schmittful, A. Challenor, D. Hanson and A. Lewis, Phys. Rev. D {\bf 88} 
063012 (2013).
\bibitem {Kroupa}      %16
P. Kroupa, M. Pawlowski and M. Milgrom, arXiv: 1301.3907.
\bibitem {WMAP}        %17
G. Hinshaw et al. arXiv: 1212.5226.
\bibitem {Alpaslan}    %18
M.Alpaslan et al. arXiv: 1401.7331.
\bibitem {Cantun}      %19
M. Cantun et al. arXiv: 1401.7886.
\bibitem {Bicep}       %20
P.A.R. Ade et al. (BICEP Collaboration), arXiv:1403.3985.
\end {thebibliography}
\end {document}